\def\etal{{\it et al.}\thinspace}
\def\ie{{\it i.e.,}\thinspace}
\def\eg{{\it e.g.,}\thinspace}
\documentclass[useAMS,usenatbib,a4paper]{mn2e}
\usepackage{graphicx}
\usepackage{geometry,natbib}
\usepackage{rotating}
\def\eq{\begin{equation}}
\def\en{\end{equation}}

\def\apj{{\it Ap.J.}\thinspace}

\def\aap{{\it A\&A}\thinspace}
\def\aaps{A\&A Suppl.}
\def\mnras{{\it MNRAS}\thinspace}
\def\P3hat{{\mathaccent 94 P}_3}

\title[An 8-hour GMRT Observation of  Pulsar B1822$-$09]{Modal Sequencing and Dynamic Emission Properties of an 8-hour GMRT Observation of  Pulsar B1822$-$09}
\author[Crystal Latham, Dipanjan Mitra, and Joanna Rankin]{Crystal Latham$^{1}$, Dipanjan Mitra$^{2}$  \& Joanna Rankin$^{1,3}$\\
$^1$Physics Department, University of Vermont, Burlington, VT 05405\thanks{Crystal.Latham@uvm.edu; Joanna.Rankin@uvm.edu} \\
$^2$National Centre for Radio Astrophysics, Ganeshkhind, Pune 411 007 India\thanks{dmitra@ncra.tifr.res.in} \\
$^3$Sterrenkundig Instituut `Anton Pannekoek', University of Amsterdam, NL-1090 GE 
}

\begin{document}
\date{Accepted 2012 month day. Received 2012 month day; in original form 2012 month day}
\pagerange{\pageref{firstpage}--\pageref{lastpage}} \pubyear{2012}
\maketitle
\label{firstpage}
\begin{abstract}
The research presented here examines an 8-hour observation of pulsar B1822$-$09, taken by the Giant Metrewave Radio Telescope. B1822$-$09 has been known to exhibit two stable emission modes, the B-mode, where the precursor (PC) `turns-on', and the Q-mode, which is defined by interpulse (IP) emission.  The results of our analysis, of this extremely long observation, have shown that B1822$-$09 exhibits at least three other emission behaviours that have not been seen before in other similar pulsars or in other observations of B1822$-$09.  These three behaviours can be described as: Q-mode emission with PC emission, B-mode emission with IP emission, and instances where both the PC and IP  are `on' when transitioning from one mode to the other.  The pulse structure has been found to be more complex than previously thought. The MP has an inner cone/core triple ({\bf T}) configuration together with a central sightline traverse. The IP is a 15\degr-wide region, that along with the MP originate from an open dipolar field. The PC emission comes from a still unknown source.  We argue that the PC emission arises within the same region as the MP, but likely comes from higher in the magnetosphere.  The Q-mode has a very clear fluctuation that occurs in both the MP and IP at 46.6-$P_1$ which is associated with drifting subpulses.  Also, we have found that the B-mode, which has previously never shown any detectable modulations at this radio frequency, has a very weak feature at 70-$P_1$.  Coincidently we find the ratio of the B-mode ``$P_3$'' of 70-$P_1$ to its Q-mode counterpart of 46.6-$P_1$ is very nearly 3/2, which seems to imply a carousel of three MP ``sparks'' in the Q-mode and two sparks in the B-mode.  The circulation times of the two modes have been found to be virtually equal at 140-$P_1$, which allows for this interpretation of the fluctuation features as ``sparks''. Overall, our analyses strongly suggest that mode changes allow information transfer between the two magnetic polar regions and contribute to global magnetospheric changes. 
\end{abstract}

\section{Introduction}
Radio pulsar B1822$-$09's fascinating characteristics have attracted the attention of many investigators who have described aspects of its modal and pulse-sequence (hereafter PS) behaviour. In this paper, we present new analyses of B1822$-$09, based on one of the longest observations ever recorded by the Giant Metrewave Radio Telescope (GMRT). Most existing studies of the pulsar's PSs had relied on observations carried out at frequencies higher than 1 GHz (\eg Fowler \& Wright 1982)---apart, that is, from our recent 325-MHz GMRT effort reported by Backus \etal\ in 2010 (hereafter Paper I) and a just-published, comprehensive low frequency analysis by Suleymanova \etal\ (2012; hereafter SLS12). In this more extensive investigation of B1822$-$09's emission characteristics, we are able to extend---and in some cases correct---Backus \etal\ 's conclusions. 

Pulsar B1822$-$09 is known to exhibit three interesting phenomena simultaneously: two ``modes'' in its average pulse profile, interpulse (IP) emission, and periodic subpulse modulation (Fowler \etal\ 1981; Morris \etal\ 1981; Fowler \& Wright 1982). B1822$-$09 switches between its `Q'uiescent and its `B'urst modes (Fowler, Morris \& Wright 1981; Gil \etal\ 1994). Some studies have reported that mode changes occur approximately every five minutes (\eg Fowler \& Wright 1982); however, our observations show that both modes can at times persist for much longer intervals. The Q-mode exhibits an IP located about half a rotation period (hereafter $P_1$) from the MP, which shows strong low-frequency modulation at some 43 $P_1$. The B mode, by contrast, exhibits a brighter and more complex MP as well as a  precursor (PC) component some 15\degr\ longitude prior to the main pulse (MP), but no detectable IP. Indeed, B-mode profiles suggest a connecting ``bridge'' between the PC and MP.  

The half-period-separated IP of B1822$-$09 suggests that the pulsar may have an orthogonal geometry, where the MP is radiated above one magnetic pole and the IP above the other (Gil \etal\ 1994).  As discussed in Paper I, earlier investigators had regarded the PC and MP as a pair of features---a conal double structure whose leading portion was present or absent in the two modes and whose centre point trailed the bright IP feature by a little less than half a period.  However, some have argued that the PC and IP are emitted by the same emission region, which reverses emission direction in its two modes (Dyks, Zhang, \& Gil 2005); whereas others have suggested a purely geometrical model for the profile of B1822$-$09 (Petrova 2008).  

\begin{figure}
\begin{center}
\includegraphics[width=\columnwidth , height = 75mm, angle = -90]{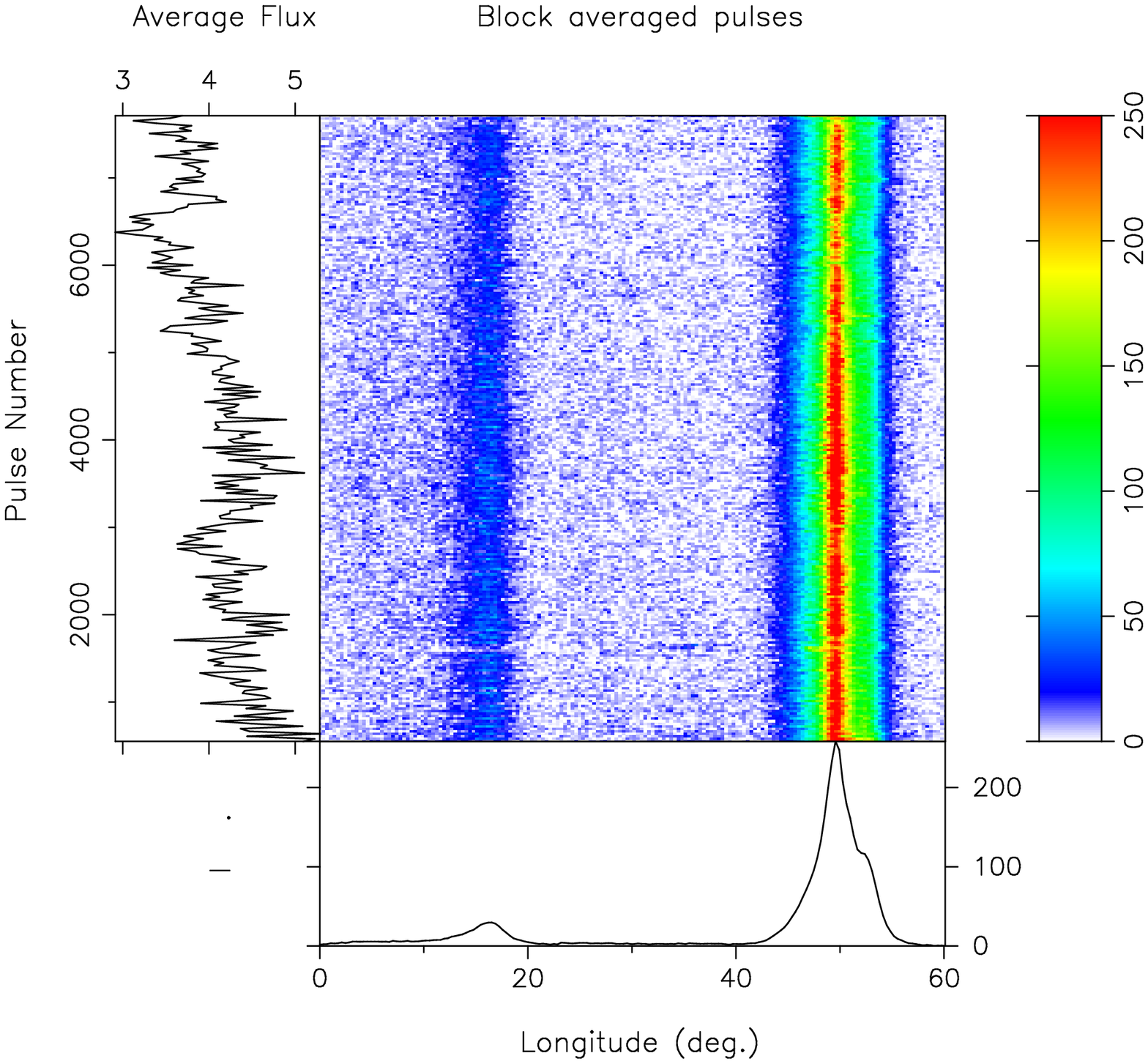}
\includegraphics[width=\columnwidth , height = 75mm, angle = -90]{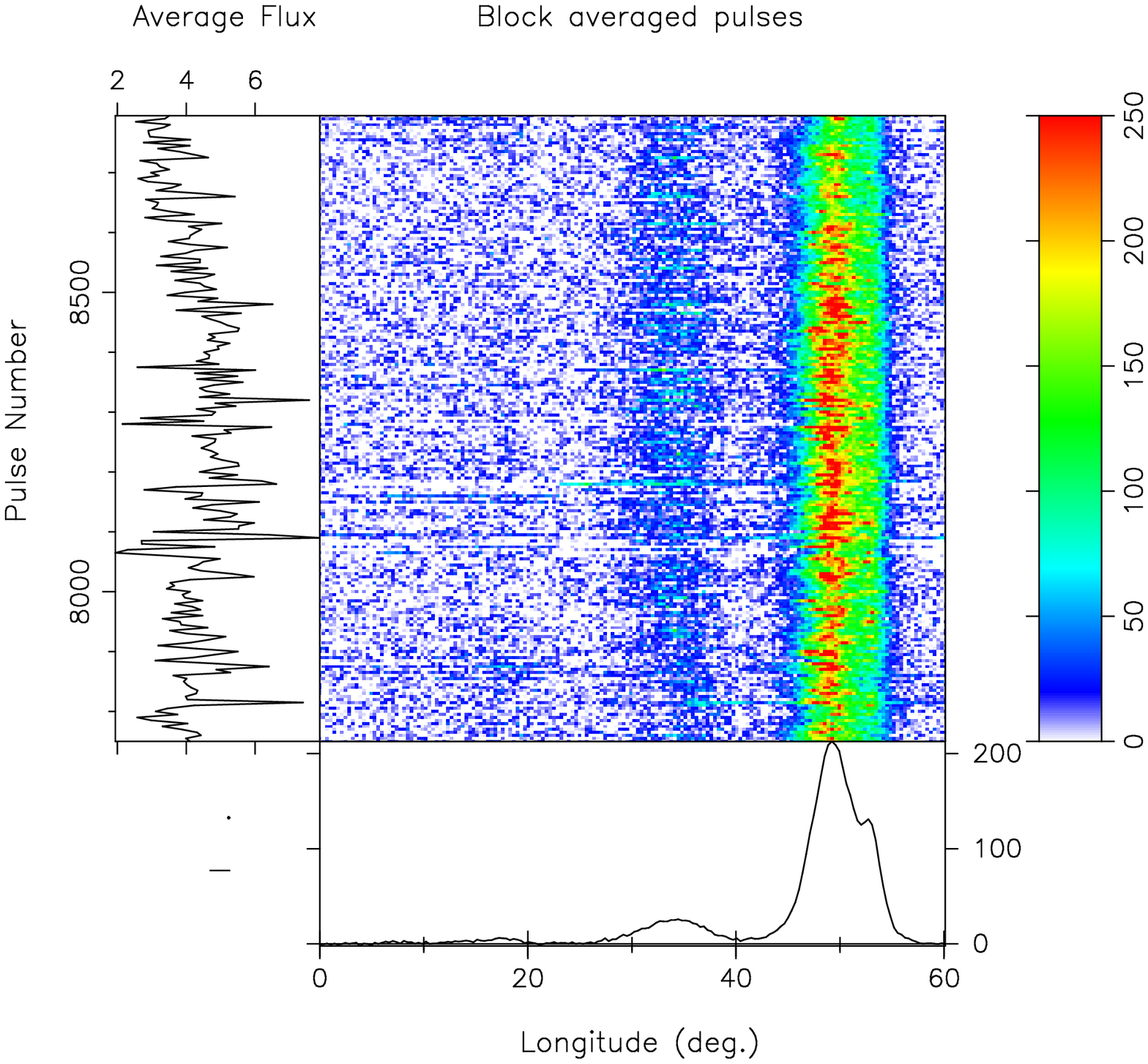}
\caption{Top: Display of the longest, 7200-pulse, Q-mode interval (pulses 20550-27750) in 29-pulse subaverages (main panel); aggregate intensity (lefthand panel); and total average profile (bottom panel). The first 10\degr\  of longitude have been removed, and a further 140\degr\ have been removed at +23\degr\ (as in Paper I), so that all three regions of the profile can be seen at once. Note the IP, relatively narrow MP, and the lack of the PC.  Bottom: Display of the longest, 1050-pulse B-mode interval (pulses 27750-28800) in five period averages as above.  Note the broader, more complex MP, the PC, and weak IP.  Here, the pulses number from 20000.}
\label{fig1}
\end{center}
\end{figure}

In Paper I, we departed from the view that the PC and MP constitute a pair.  We found that the MP has three partially merged features, and that these have the expected angular dimensions for an inner-cone/core triple ({\bf T}) profile, viewed in a nearly orthogonal sightline geometry.  Our sensitive observations also showed that the IP has what is probably a conal double ({\bf D}) structure, rather than the single feature seen previously---and that the midpoints of the IP and MP are separated by almost exactly 180\degr. In summary, the evidence appears very strong indeed that the IP and MP are emitted above the star's two magnetic poles, though this understanding leaves the PC without any ready interpretation within the core/double-cone model.  

Previous studies of B1822$-$09 give the impression that the two modes each appear roughly half of the time. However, one of the PSs described in Paper I consisted of an unusually long (2106-pulse) Q-mode observation; whereas in the other the longest B-mode interval was only 255 pulses. In this analysis, we have acquired an 8-hour, 37399-pulse GMRT observation, of which approximately 10\% or 3740 pulses were omitted from analysis owing either to RFI or telescope phasing.  Investigation of the remaining 33000+ pulses have provided a number of new opportunities. Indeed, the observation includes the longest Q-mode PS ever recorded, a total of 7200 pulses as well as a B-mode interval some 1050 pulses in length, four times longer than was available to Backus \etal .

Our unprecedentedly long meter-wavelength observation provides an opportunity to ask new questions about the character of B1822--09's emission. In terms of sequencing, we find brief instances when the modes appear to be in transition as well as instances where the PC and IP appear simultaneously. We have also found that the PC, IP, and MP have multiple features within their profile structure, which can be intercompared and interpreted. On close inspection, the MP intensity varies considerably, and at times almost nulls, such that both the IP and PC are undetectable. The several long unimodal PSs facilitate detailed fluctuation-spectral analyses, and in particular we find a long period Q-mode modulation feature common to both the MP and IP for the first time. Finally, we use the results of the analyses to assess current models pertaining to the emission geometry and characteristics of this pulsar.

\begin{figure}
\begin{center}
\includegraphics[width = .9\columnwidth , height = 75 mm, angle = -90]{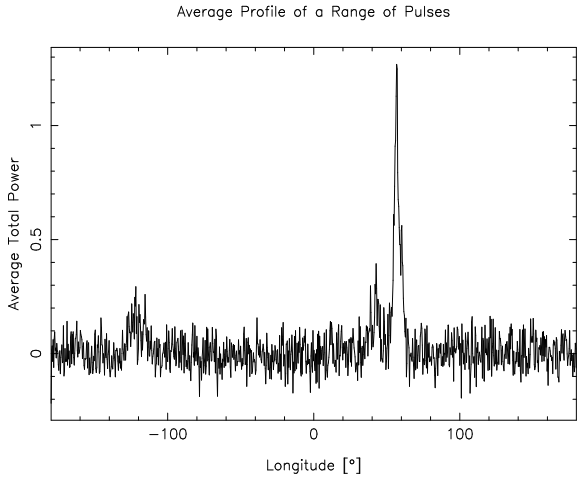}
\caption{Four-pulse average profile (20548-20551) showing the transition region to the long Q-mode apparition in Fig.~\ref{fig1} (upper). Here we see a good example of simultaneous IP and PC illumination.}
\label{fig2n}
\end{center}
\end{figure}

\begin{figure}
\begin{center}
\includegraphics[width=\columnwidth , angle = 0]{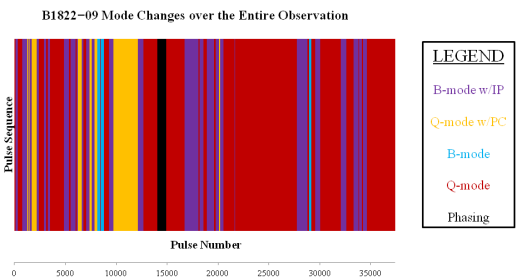}
\caption{Modal sequencing in the 8-hour (37399 pulses) GMRT observation. The Q-mode occurs more often and in longer intervals that the B-mode. Also, the IP `sputters on' in B-mode intervals (Bw/IP) and the PC in Q-mode sections (Qw/PC).  Alternating intervals of Q and Bw/IP are also noticeable (seen as red and purple patterns). The black region corresponds to the time section used for phasing the antenna.}
\label{fig3n}
\end{center}
\end{figure}

In \S 2, we describe the GMRT observation of B1822$-$09, and in \S 3, we discuss the modal profile characteristics of the pulsar, including its pulse structure, MP emission, and the pulse-intensity variations with time. \S 4 addresses the pulsar's moding behaviours; while in \S 5 we present the results of our fluctuation-spectral analyses. In \S 6, we summarize our main findings and their implications, as well as discuss how these results should be interpreted in the context of the various theoretical interpretations of B1822--09's emission properties.
\maketitle

\section{Observation}
The 8-hour, 325-MHz observation of B1822$-$09 provides the basis for our analyses in this paper. The observations were carried out at the Giant Metrewave Radio Telescope (GMRT) near Pune, India. The GMRT is a multi-element aperture-synthesis telescope (Swarup \etal\ 1991) consisting of 30 antennas distributed over a 25-km diameter area which can be configured as a single dish both in coherent and incoherent array modes of operation. For these observations we used the coherent (or more commonly known as phased-array) mode of the GMRT. A bandwidth of 16 MHz was used (whole band running from 325 to 341 MHz) and the data was recorded using the GMRT pulsar machine at a time resolution of 0.512 milliseconds. The observations used 20 GMRT antennas (14 cental square and 2 each in the east west and south arm) for phasing. Usually the calibrator source used for phasing lies several degrees away from the pulsar and one needs to rephase the array every 1.5 hours which causes missing pulsar-less time sections in the data. However, for this observation the phase calibrator used was 1822--096 lying in the same primary beam of the telescope, and hence phasing could be done while the pulsar was always on the beam. This way we were able to record 37399 continuous pulses almost without breaking the pulsar data stream. Only between pulse number 14016 to 14910 due to interference the telescope was slewed to a nearby calibratrator source for checking the phasing. 

After dedispersion most of the data was of good quality. However, in sections there were large baseline variations due to broadband interference. This was mostly mitigated by running mean-subtraction of the baselines as well as flattening of the baseline for every single pulse. No significant power line RFI was present in the dedispersed data stream.

\begin{figure}[th]
\begin{center}
\includegraphics[width=\columnwidth , height = 75mm, angle = -90]{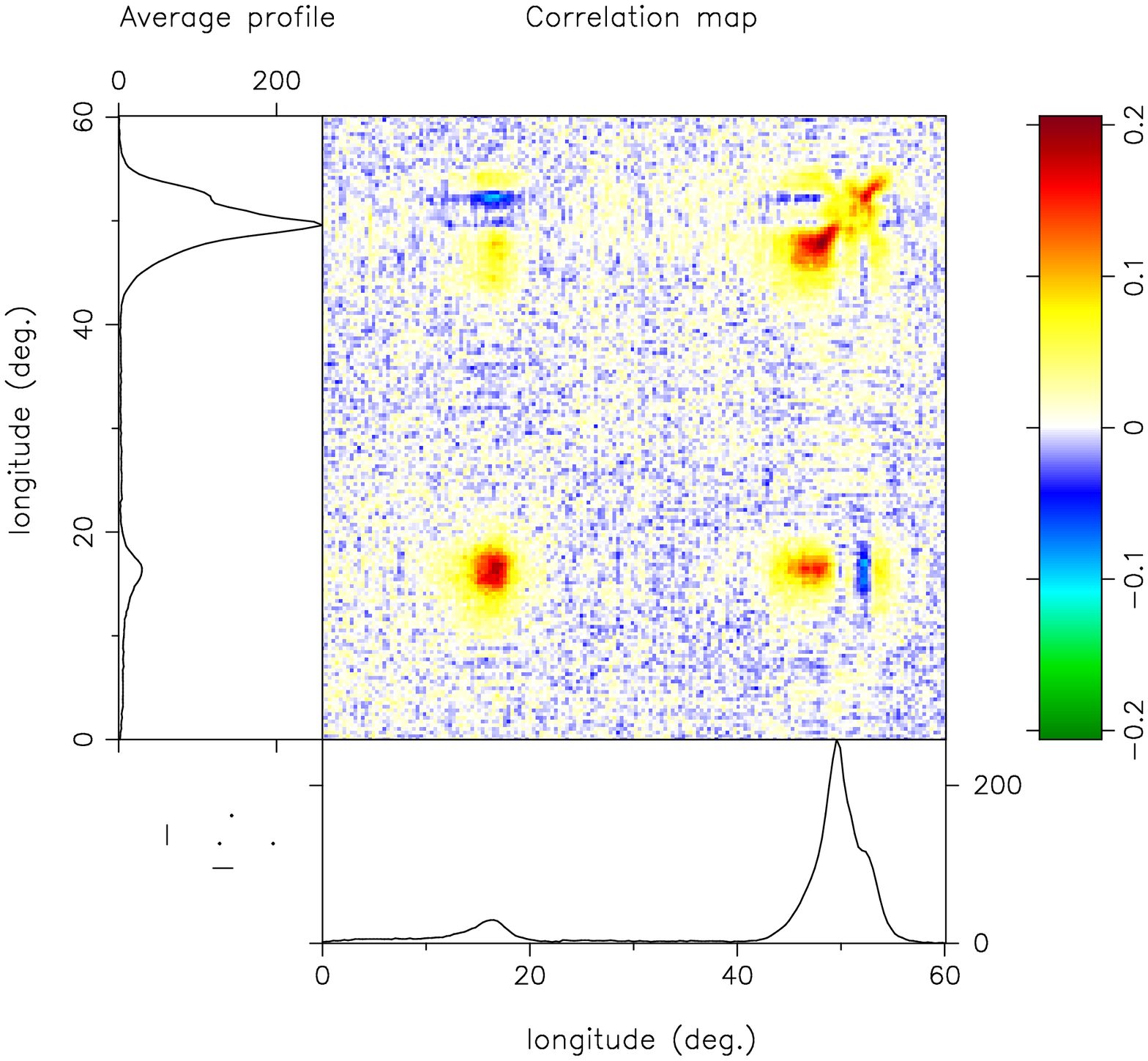}
\includegraphics[width=\columnwidth , height = 75mm, angle = -90]{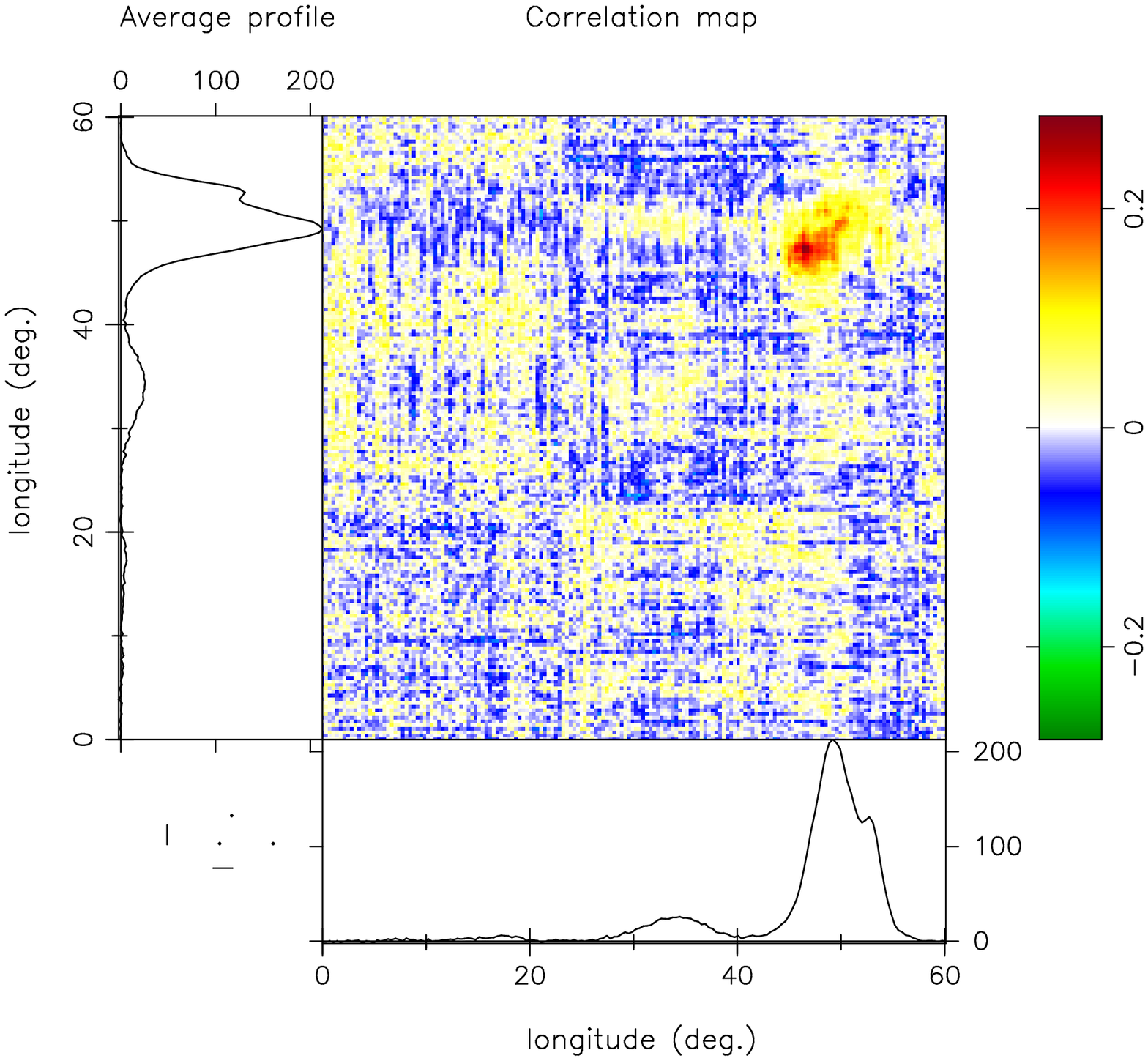}
\caption{Longitude-longitude correlation displays at one-pulse lag corresponding to the long Q- (upper) and B-mode (lower) intervals in Fig.~\ref{fig1}. In the Q-mode, the MP correlates substantially both with itself and the IP. Indeed, the leading, centre, and trailing parts of the MP exhibit different amounts and types of correlation, demonstrating its triple character. Only the IP peak correlates, but the remainder is probably too weak for this analysis. However, the B-mode self-correlation in the MP, its cross-correlation with the PC, and the PC's self-correlation are weak to negligible.  The contrast with the Q-mode could not be more marked.  The three-$\sigma$ errors on the correlations here is 0.035 and 0.093, respectively.}
\label{fig4n}
\end{center}
\end{figure}

\section{Modal Behavior}
Pulsar B1822$-$09's modal effects are among of its most fascinating and important behaviours.  Following the convention of earlier investigators, we discuss its modes in terms of its ``quiescent'' Q-  and ``bright'' B-modes. The Q-mode is by far the most usual mode; it persists through 60\% of our observation. The longest (very nearly) continuous Q-mode apparition is 7201 pulses long, which is shown in the upper portion of Figure~\ref{fig1}. By contrast, the B-mode persists for only 30\% of the observation, and its longest continuous apparition is 1151 pulses in length, shown in the lower portion of Fig.~\ref{fig1}. These two modes define the emission from B1822$-$09, and in our observation we have discovered that these two modes occur in parallel to create three other behaviours: Q-mode with PC activity (Qw/PC), B-mode with IP activity (Bw/IP), and both together at modal transitions.

\subsection{Parallel Moding}
Looking closely at the bottom B-mode display of Fig.~\ref{fig1}, an obvious low level of IP emission can be discerned at the beginning of this long sequence. Nearly all intervals of consistent B-mode emission, some 93\%, also contain instances of IP emission. Similarly, in the upper Q-mode display, an interval can be seen around pulse 21620 where the B-mode appears to flicker `on' for a few pulses. The former is just one clear instance wherein the usually mutually exclusive PC and IP are found to occur simultaneously. At other times both the IP and PC seem to be `off' briefly---but this might be expected because in the Q-mode the IP is cyclical. Overall, we have found it challenging to fully characterize as long an observation as the present one at the single-pulse level; however, it is very clear that the two modes are not as ``pure'' as previously thought. Such instances tend to occur during transitions between the modes (see for \eg Fig.~\ref{fig2n}), though the above examples occur well into long modal intervals.  

Moding ``in parallel''---that is, simultaneous IP and PC activity---is a phenomenon that had not previously been identified in pulsar B1822$-$09. In the foregoing example, IP emission occurred within an otherwise ordinary B-mode interval, so one might dub the effect B-mode with IP (or Bw/IP), and indeed this the most common form of parallel moding.  About 29\% of the pulses in our observation have been classified as the Bw/IP mode. Though not nearly as common, the Q-mode with PC (Qw/PC) also occurs throughout our observation in about 11\% of the pulses. The Qw/PC intervals are identified by PC emission within an otherwise normal Q-mode PS. Finally, a third type of parallel moding sometimes occurs when B1822--09 transitions from B to Q or vice versa.  

\subsection{Modal Sequencing}
B1822$-$09 is known for its unique modal sequencing, and with our extremely long 
observation we were able to see changes and patterns of moding throughout. 
Figure~\ref{fig3n} gives a diagram summarizing the modal configuration over the 
entire observation.  Noisy areas of the observation have been integrated into the 
respective modal sequence in which it occurs to simplify the presentation. The most 
common pattern is for the modes to alternate from Q to Bw/IP and back to Q.  In this 
specific pattern, the transition typically has a duration of a few pulses.  A few instances 
of Bw/IP to Qw/PC and back to Bw/IP were also identified.  Interestingly, the two longest 
intervals of relatively pure Q- and B-mode occur together, right after one another. 
Previous studies had found that mode switching occurs approximately every five 
minutes (\eg Fowler \& Wright 1982); however, here we have found that the average 
time between mode changes is about 7.6 minutes, or a pulse length of 593 pulses.

\section{Average Profile Structure}
Pulsar B1822$-$09's MP, PC, and IP have long been studied, but only recently 
has a full and consistent picture of these structures emerged.  Here, we have the 
benefit of long continuous B- and Q-mode intervals, which we find confirm and 
extend the viewpoints developed in Paper I.  Figure~\ref{fig1} (upper display) 
shows the longest Q-mode interval, 7201 pulses, in 29 period averages, spanning 
pulses 20550-27750; whereas; Fig.~\ref{fig1} (lower) shows the longest B-mode 
sequence, from pulse 27750 to 28800, in five period averages.  We can see clearly 
again here that the MP amplitude is similar in the two modes, so that the B-mode is 
largely brighter due to its increased width.  

\begin{figure}
\begin{center}
\includegraphics[width=\columnwidth , height = 80mm, angle = -90]{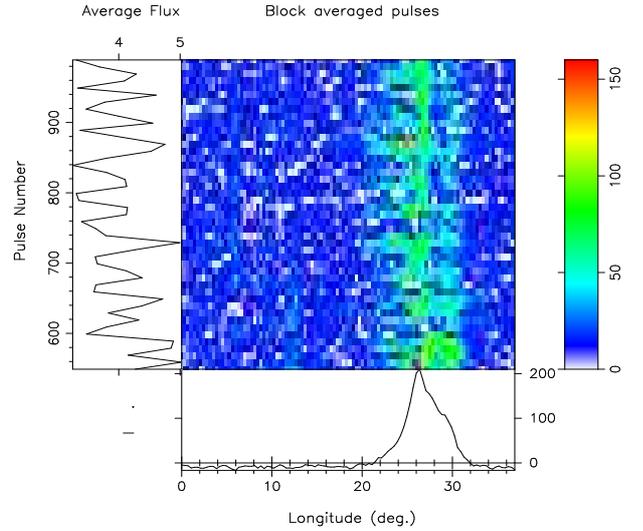}
\caption{Beginning of the long Q-mode PS in Fig.~\ref{fig1} in 10-pulse averages.  Here 
we see clearly that emission associated with the trailing part of the profile decreases.}
\label{fig5n}
\end{center}
\end{figure}

\begin{figure}
\begin{center}
\includegraphics[width=\columnwidth , height = 80mm, angle = -90]{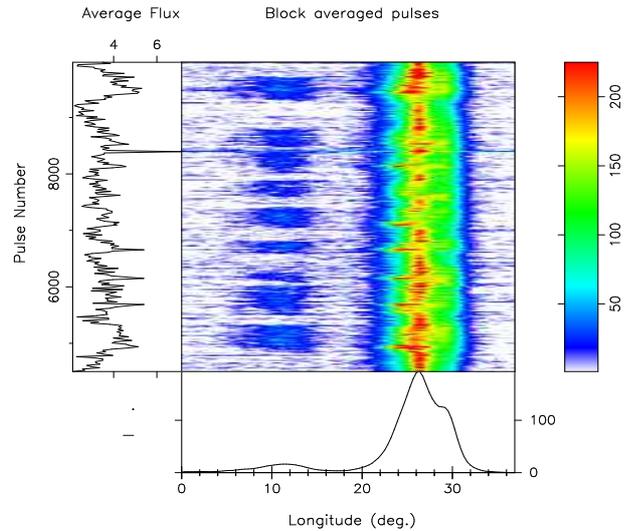}
\caption{B-mode episodes during pulses 4501-10000 in 22-pulse averages.  Note the 
strong emission on the leading edge of the MP that coincides with the beginning of each 
B-mode interval.  Also, the average flux (left-hand panel) decreases markedly during each 
apparition.  The strong response near pulse 8400 is RFI. }
\label{fig6}
\end{center}
\end{figure}

\subsection{Modal Profile Structure}
Paper I argued that the MP was comprised of three poorly resolved features, and 
that it reflected a nearly orthogonal (magnetic latitude $\alpha$ about 90\degr) 
inner cone/core triple ({\bf T}) configuration together with a central sightline traverse 
(impact angle $\beta$ near 0\degr).  Using the long B- and Q-mode intervals in this 
observation, we have computed peak-occurrence histograms (not shown) for the MP 
in both modes.  Unsurprisingly, these show sharp peaks at sample 674 (of 1024), 
but also the broad leading and trailing pedestals indicative of a triple structure.  The 
mixed-mode structure of the MP is remarkably similar within the two octaves below 
1.6 GHz (\eg Gould \& Lyne 1998), and SLS12 show that this 
basic structure persists with some broadening down to 62 MHz.  

Figure~\ref{fig4n} (upper) exhibits the pulsar's Q-mode profile structure using correlation 
maps:  leading and trailing regions of the MP correlate both with each other and 
the IP; whereas, the central region of the MP shows correlation only with itself.   
The corresponding B-mode correlations are then shown in Fig.~\ref{fig4n} (lower), and 
the difference could not be more marked.  Here we see negligible correlation both 
between the MP regions as well as with the PC.  

Further, Paper I showed that the IP extended 10\degr\  or so earlier than the single 
trailing peak seen at higher frequencies.  In our observation that peak occurs at 
sample 179, so that the interval between the MP and IP peaks is indeed the often 
quoted 186\degr.  However, the IP mid-point lies 6-7\degr\ earlier, so that the MP 
and IP profile centres are almost exactly half a period apart.  Had there been any 
lingering doubt about this interpretation, SLS12's remarkable 62-MHz 
IP detection---showing a symmetrical dual-lobed IP (their fig. 6) resolves it.  They 
also show that its 62-MHz midpoint follows the MP peak by almost exactly 180\degr, 
which together with (corrected) high frequency measurements (Hankins \& Fowler 
1986) show that the MP-IP separation is constant within small uncertainties over 
the entire range of available observations. Finally, SLS12's fig. 7 shows that the MP 
and IP profile widths are comparable and escalate comparably in a conal fashion.

\begin{figure}
\begin{center}
\includegraphics[width=\columnwidth , height = 80mm, angle = -90]{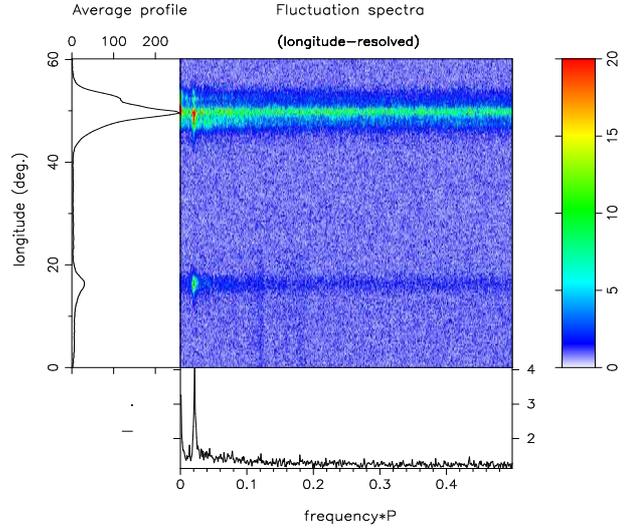}
\caption{Longitude-resolved (hereafter LRF) spectra for the longest Q-mode interval in 
Fig.~\ref{fig1} (upper; pulses 20550-27750), using a 1024-point FFT.  The MP is at the 
top and the IP at the bottom of the left-hand panel.  The strong feature at about 0.021 
cycles/period (hereafter c/$P_1$) modulates both the MP and the IP and is coherent 
here but not using a 2048-length FFT; thus, it corresponds to a $P_3$ of 46.6 $P_1$ 
with an uncertainty of about $\pm$0.9 $P_1$. The nature of $P_3$ will be discussed
and interpreted in \S 5.5.} 
\label{fig7}
\end{center}
\end{figure}

\begin{figure}
\begin{center}
\includegraphics[width=\columnwidth , height = 75mm, angle = -90]{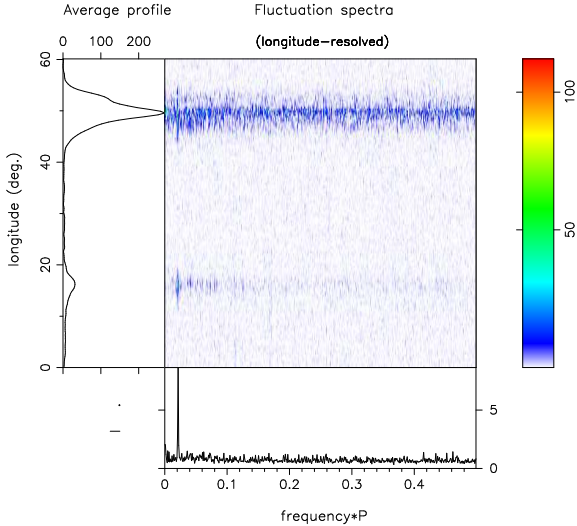}
\caption{LRF spectra as in Fig.~\ref{fig7} for a highly modulated region near the 
centre of the long Q-mode apparition (pulses 23622-24645) using a 1024-length 
FFT.}
\label{fig8}
\end{center}
\end{figure}

Longitude-longitude correlation plots confined to the IP (not shown) further 
elucidate its structure.  Only the region around the peak is correlated with itself 
at either zero or 1-pulse delay.  Similar displays computed for lags of 23- and 
46/47-$P_1$ show comparable levels of negative and positive correlation, 
respectively---and again no perceptible correlation with the early part of the IP.

The foregoing understanding of the MP and IP then leaves the PC without any 
plausible interpretation within the double-cone/core model.  As emphasized in 
Paper I, its large fractional polarization and shallow polarization-position-angle 
(hereafter PPA) contrast sharply with the MP and IP.  Its Gaussian-shaped, single 
form peaks in our observation at sample 631 with a half-power width of 7.0\degr, 
thus leading the MP peak by 15.1\degr---and reference to high frequency profiles 
(\eg Gould \& Lyne's at 1.6 GHz) shows a virtually identical form and spacing.  The 
PC's RF spectrum is substantially flatter than that of the other components as can 
be seen in Gould \& Lyne's profiles, and this trend continues at low frequencies 
where only hints of the PC are seen in SLS12's decimetre profiles.  
We also attempted to explore the PC's structure using correlation analyses (\eg 
Fig.~\ref{fig4n} (lower)), but the feature self-correlates only narrowly and shows negligible 
pulse-to-pulse correlation.

\subsection{Secular Changes in Pulse Intensity}
Paper I noted a secular decrease in the amplitude of the two main component peaks 
during the early part of a Q-mode sequence (their fig. 14), and this effect was interesting 
because it seemed to mirror what had been seen earlier in pulsar B0943+10 following 
its B-mode onset.  Here, we have more opportunities to examine the intensity behaviour 
during both modal episodes of the profile.  Figure~\ref{fig5n} shows the beginning of the 
long Q-mode sequence of Fig.~\ref{fig1} (upper) in 10-pulse averages, and we see clearly 
that the intensity of the trailing component diminishes over the first 100 or so pulses.  
Indeed, given that in general the B-mode PSs are more intense that those of the 
Q-mode---and that such intensity changes occur at the modal transition---we might 
regard this excess power in the trailing component as ``left over'' from the preceding B-mode.  
Systematic intensity changes are also seen in the B-mode sequences.  Figure~\ref{fig6} 
shows a series of short B-mode episodes between pulses 4501 and 10000.  First, a 
dramatic intensity increase in the early part of the MP coincides with the beginning of 
each B-mode episode, and second, the overall intensity (left-hand panel) is seen to 
fall off during most of the B-mode intervals.

\section{Fluctuation-spectral Analyses}
Pulsar B1822--09's prominent 43-$P_1$ Q-mode fluctuation feature is well known from 
earlier studies (\eg Paper I).  It modulates the IP peak as well as both leading and trailing 
regions of the MP.  However, a fluctuation cycle this long is difficult to measure accurately, 
so little is known about its precise frequency, possible variations, or behaviour near modal 
transitions.  Earlier studies using 256-length FFTs entail a half-bin error or $\pm$3.5 $P_1$, 
so the 43-$P_1$ value must be considered a nominal one.

\subsection{Q-mode Modulation Frequency}
Here, the very long observation and unprecedentedly long Q-mode intervals provide a 
possibility of improved precision and better answers to the above question.  In particular, 
Figure~\ref{fig7} shows longitude-resolved fluctuation (hereafter LRF) spectra for the 
longest Q-mode apparition in Fig.~\ref{fig1} (upper).  Using a 1024-length FFT, all the 
fluctuation power falls in a single bin; whereas, in an FFT of twice this length the power 
is divided about equally.  Thus the modulation period and uncertainty can be given as 
46.55$\pm$0.88 $P_1$.  

\begin{figure}
\begin{center}
\includegraphics[width=\columnwidth , height = 75mm, angle = -90]{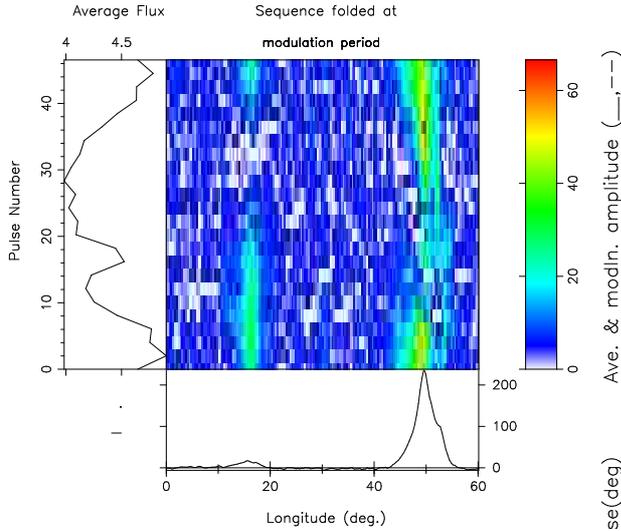}
\caption{Folded PS of the highly modulated Q-mode region in Fig.~\ref{fig8} with 
the non-fluctuating ``base'' (shown in the bottom panel) removed.  We see here 
that the IP remains stationery in longitude throughout the modulation cycle, 
whereas the MP fluctuates in a highly ordered and ``drifting'' manner.}
\label{fig9}
\end{center}
\end{figure}

\begin{figure}
\begin{center}
\includegraphics[height = 75mm, angle = -90]{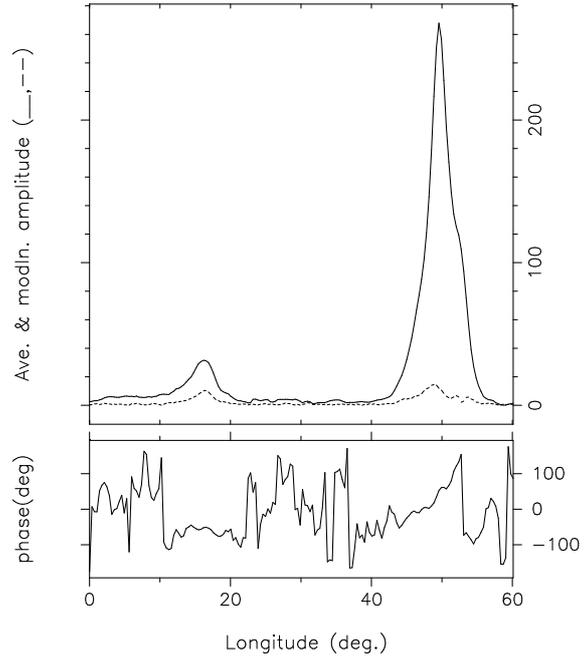}
\caption{Modulation phase as a function of longitude for the highly modulated Q-mode 
region in Fig.~\ref{fig8}.}
\label{fig10n}
\end{center}
\end{figure}

Over the duration of the longest Q-mode interval, we found that the strength of the 
modulation varied greatly, but no deviation from this period could be discerned.  For 
instance, Figure~\ref{fig8} gives LRF spectra for the most highly modulated region 
near the centre of the Q-mode apparition (pulses 23622-24645).  Here again, all 
the modulation power is in one bin, so its period and error is as above.  Overall, 
the strength of this modulation seemed to grow stronger during the first half of the 
apparition and diminish in the latter half.  Near the end, just before the B-mode 
onset (\eg pulses 26727-27750), the feature was very weak and the fluctuation 
power divided between the two adjacent bins in a 1024-length FFT, but the feature 
period was indistinguishable from the above, within its larger error.   

Our 8-hour observation provided a number of other opportunities to measure the 
Q-mode modulation frequency, but not for as long a continuous sequence as the  
one above.  Other extended such intervals were found roughly between pulses  
3500-4800, 10300-12100, 12700-14000, 14900-16700, 30050-31250 and after 34500.  
In each of these, the modulation feature was easily seen in both the IP and MP, but 
was less coherent than in the most stable intervals above.  Usually the fluctuation 
power was divided between two bins in a 512-length FFT; however, its period was
always comparable to the value above and within its error.

\subsection{Character of the Q-mode Modulation}
We have seen above that both the IP and MP are modulated at the same period 
and that their fluctuations are highly correlated (\eg Fig.~\ref{fig4n}).  However, it is 
important to know in detail just where and how the intensities vary.  The display of 
Figure~\ref{fig9} then shows a modulation-folded PS of the most coherent region 
of the longest Q-mode apparition in Fig.~\ref{fig8}.  Here the non-fluctuating ``base'' 
(bottom panel) has been removed, so only that part of the power that fluctuates is 
shown.  Clearly, the IP fluctuates in a longitude-independent manner; whereas, the 
MP exhibits a complex behaviour over the course of the modulation cycle.  

Taking the IP first for a more detailed analysis, we produced a modfold similar to 
Fig.~\ref{fig9} restricted to the IP by itself (not shown), and we saw again that the modulation 
is longitude independent and upwards of 50\% modulated (note the range of values 
in the left-hand panel or compare the ``base'' amplitude in the lower panel with that 
of the colour bar).  It will not then be surprising that an harmonic-resolved fluctuation (HRF) 
spectrum corresponding to the LRF in Fig.~\ref{fig8} (not shown)---restricted to the 
IP---shows the dual features indicative of amplitude modulation.  Or what is the same, 
a modulation-phase analysis (see Fig.~\ref{fig10n}) for the same interval shows it constant 
over the IP ``peak'' region (and the fluctuation power in the leading IP region is so 
small that the phase is meaningless).  While these results for the ``coherent'' interval 
of Fig.~\ref{fig8} are especially sensitive, we have applied the same analysis to most 
other Q-mode regions in the observation with virtually identical results.  

\begin{figure}
\begin{center}
\includegraphics[width=\columnwidth , height = 75mm, angle = -90]{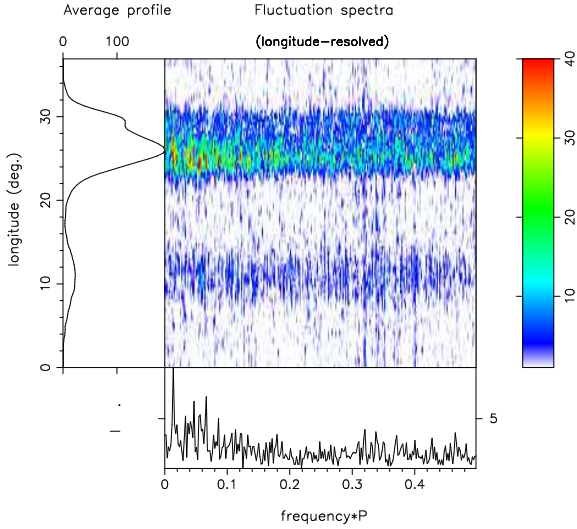}
\caption{The LRF spectra of the longest B-mode sequence, from pulse 27751-28800, 
using a 512 point FFT as in Fig.~\ref{fig7}.  A strong feature at a period of 70$\pm$3 
$P_1$ is evident as well as several more rapid periodicities.  The colour-scale is 
truncated to show the weak features more clearly.  During this interval the pulsar flux 
varies substantially (see Fig.~\ref{fig1}), so it was necessary to correct for this possible 
scintillation.}
\label{fig11}
\end{center}
\end{figure}

\begin{figure}
\begin{center}
\includegraphics[width=\columnwidth , height = 80mm, angle = -90]{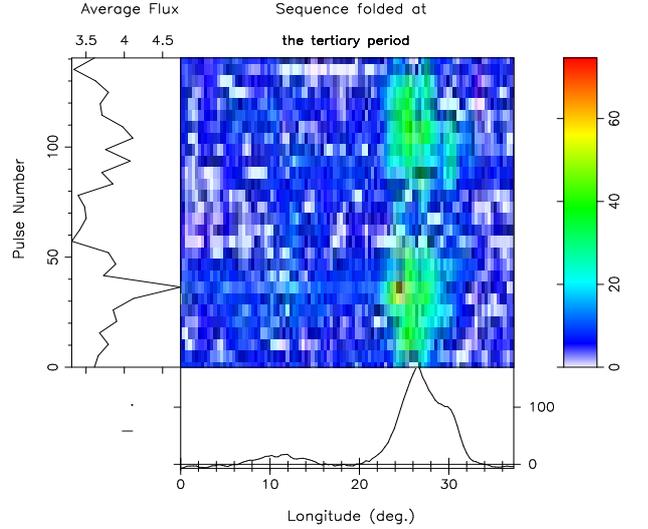}
\caption{Folded PS of the longest B-mode episode, from pulse 27751-28800 as in 
Fig.~\ref{fig11}.  A folding period of twice the exact 70-$P_1$ response was used.  
All parts of the MP exhibit the 70-$P_1$ fluctuation, and there is a suggestion that 
it involves the PC as well.  In fact, this modfold suggests that the PC fluctuations  
are stronger every other 70-$P_1$ cycle.}
\label{fig12}
\end{center}
\end{figure}

\begin{figure}
\begin{center}
\includegraphics[width=\columnwidth , height = 80mm, angle = -90]{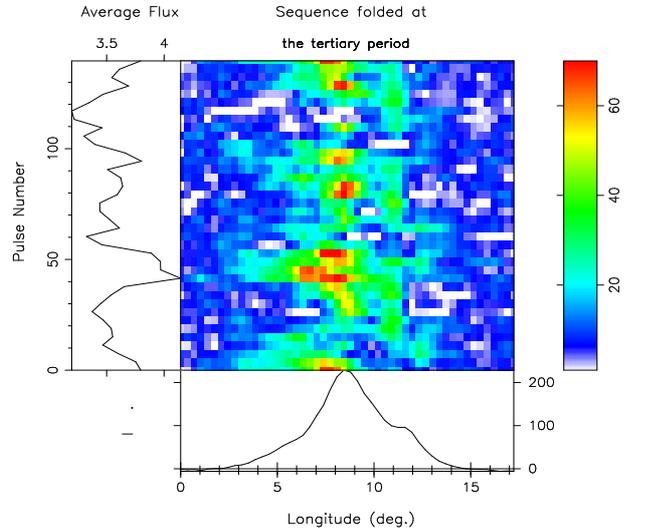}
\caption{Folded PS of the Q-mode MP alone as in Fig.~\ref{fig9}, but here at three times 
the 46.6-$P_1$ cycle.  The corrugation is substantially larger, suggesting that this 
longer cycle is more fundamental.}
\label{fig13}
\end{center}
\end{figure}

\begin{figure}
\begin{center}
\includegraphics[width=\columnwidth , height = 75mm, angle = -90]{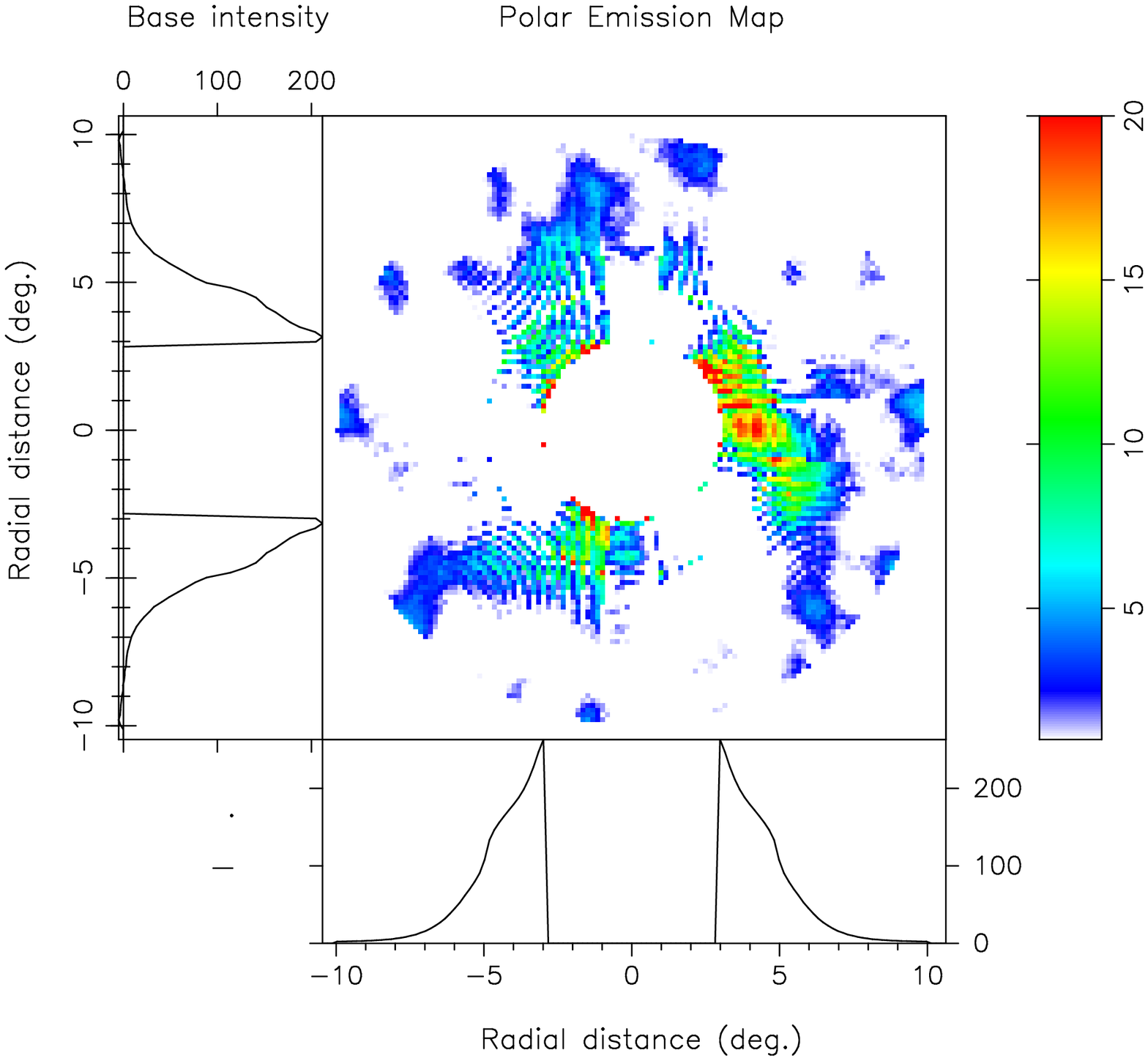}
\caption{Polar map of the MP sub-beam structure in the PS of Figs.~\ref{fig9} and \ref{fig13}.  
A nearly orthogonal geometry where an $\alpha$, $\beta$ of 85\degr\ and 4\degr, respectively was used. 
It is important to note that the display is not sensitive to the precise values as long as $\beta$ 
remains less than about 5\degr.}
\label{fig14}
\end{center}
\end{figure}

Now examining the MP in much the same manner, we computed a modfold similar 
to Fig.~\ref{fig9} restricted to the MP region by itself (not shown).  Here we saw that 
every part of the MP profile fluctuates in a dynamic and longitude-dependent manner.  
The central region near the peak fluctuates least and in the most stationary manner, 
as can be inferred from the aggregate-flux variations in the left-hand panel.  Also, 
power near the trailing edge of the profile seems to participate in the cycle at a fixed 
longitude.  The bulk of the fluctuating power, however, moves through the profile from 
the trailing to the leading side in a kind of ``drift'' over roughly half the cycle.  This is
clearly seen in an HRF (not shown) similar to the LRF of Fig.~\ref{fig7}---but for 
the MP region on its own---where the fluctuating MP power is seen to represent an 
almost pure phase modulation.  

Again, these dynamic characteristics of the MP modulation are most clearly seen in 
the highly ``coherent'' region of the longest Q-mode apparition.  However, we have 
carried out similar analyses for each of the shorter Q-mode intervals, and they all 
seem to behave similarly---that is:  a) the peak region exhibits a largely stationary, 
roughly 10\% modulation; b) a region of stationary fluctuation can be discerned on 
the trailing edge of the profile just lagging the secondary peak; and c) most of the 
fluctuation power moves through the profile from the trailing to the leading side 
during the strong half of the cycle.  

Finally, we have seen above that the IP and MP fluctuations are correlated, but it 
was not fully clear just how.  Figure~\ref{fig10n} then gives a partial answer to this 
question.  This modulation-phase diagram includes both the IP and MP, and we see 
immediately that the IP peak-region phase is stationary whereas the modulation 
phase under the MP increases with a sharp ramp under most of the profile.  Of 
course, we are dealing here with the phase of a roughly 46-$P_1$ fluctuation, so 
high temporal precision is impossible.  Nonetheless, it is remarkable that the IP is 
highly correlated (positively) with the leading edge of the MP as well as being both 
negatively and positively correlated with regions on the MP trailing edge.  And this 
is exactly what appears to be indicated in the 1-period-delay correlation map of 
Fig.~\ref{fig4n}.

\subsection{B-mode Modulation Frequency}
Figure~\ref{fig11} gives LRF spectra for the longest B-mode sequence in our 
observations seen earlier in Fig.~\ref{fig1} (lower).  A significant response at a period 
of about 70$\pm$3 $P_1$ is evident as well as several weaker features.  The 
colour scale was compressed to show the weaker responses more clearly, 
especially the weak modulation under the PC at about 0.06 c/$P_1$.  The 
70-$P_1$ feature was not ``coherent'' in an FFT of length 1024, but was so 
using 512; and while strong in the two halves of the PS, it was substantially 
stronger in the first half.  Gil \etal\ (1994) has reported detecting an 11-$P_1$ 
feature at higher radio frequencies, but we see no evidence of this in our observation.  
In only one other section of our observation was another long B-mode apparition 
available, this in the PS from 16701 and 17724; and here no significant periodicity 
was detected.

\subsection{Character of the B-mode Modulation}
A folded PS of the longest B-mode interval at 70-$P_1$ showed strong corrugation 
corresponding to an overall modulation depth of about 20\%.  All parts of the MP 
were seen to fluctuate as well as a hint that the PC also participated to some degree.  
Interestingly, a folded PS at twice this period, shown in Figure~\ref{fig12}, exhibited a 
somewhat larger depth and a suggestion that the PC fluctuations are stronger on 
alternate cycles.  Clearly, we have no other comparable B-mode interval to compare, 
and the less stable modulation feature in the B-mode prevented our pursuing the 
analysis further.

\subsection{Evidence of Carousel Circulation}
Interestingly, the corrugation in the Q-mode MP folded PS increases significantly at 
three times the 46.6-$P_1$ feature period.  Figure~\ref{fig13} shows the result, and 
no other small multiplier of the feature period increased the corrugation noticeably.  
This suggests, that the 46.6-$P_1$ feature should be interpreted as a $P_3$, not a 
circulation time (hereafter CT).  

In fact, were B1822--09's CT found to be as short as 46-$P_1$, interpretation would 
be awkward, because the pulsar's expected circulation time, according to Ruderman 
\& Sutherland's (1975; hereafter RS) relation is at least 43-$P_1$, were it as 
slow in relation to this prediction as most other pulsars with known CTs, it could easily 
be as long as 250-$P_1$.  This is because the pulsar's magnetic field is larger and the
period shorter, than for instance, B0943+10.  

Therefore, it is interesting to explore whether 3x46.6-$P_1$ or 140-$P_1$ might 
represent the star's Q-mode CT.  By way of exploration we computed a polar map 
using this putative CT for the highly modulated sub-PS of Fig.~\ref{fig8} and the 
result is shown in Figure~\ref{fig14}.  As expected, three broad subbeams can be 
seen with one somewhat brighter than the others.  Given, however, that this map 
simply reflects the modfold of Fig.~\ref{fig13}, it proves nothing by itself.  Rather, 
it provides a means of visualizing how the various correlations computed earlier 
might be interpreted.  Fig.~\ref{fig10n} shows that one of the beams, the one on the 
MP leading edge---correlates in phase with the IP; however, it also indicates that 
there will be two trailing regions, an inside one that will not correlate in phase and 
a later one that will.  Is this not exactly what we see in the delay-1 correlation map 
of Fig.~\ref{fig4n} (upper).  Moreover, while the quality of the corresponding B-mode 
map (lower) is poor, there is no region of negative correlation---as would be 
expected if there are two rather than three ``beamlets''.

Finally, we note that the periodicities identified in the respective Q- and B-modes 
appear to be commensurate.  The ratio of the B-mode ``$P_3$'' of 70-$P_1$ to its 
Q-mode counterpart of 46.6-$P_1$ is very nearly 3/2.  But even more interesting 
is the appearance that the CTs of the two modes are virtually equal at 140-$P_1$.  
If this interpretation is correct, it is the longest ($\sim$ 107.6 sec) CT yet identified 
in any pulsar.

\section{Summary and Conclusions}

Pulsar B1822$-$09 belongs to a small group of pulsars that have interpulse (IP) 
emission and is unique in that it also exhibits prominent changes of mode.  Moreover, 
its strongly anticorrelated IP and precursor (PC) emission have fascinated pulsar 
investigators and inspired a rich literature.  Key, then, to every question are the 
sources---that is, the magnetospheric geometry of the three corresponding regions 
of emission.

The recent sensitive, P-band PS observations, however, both above and in Paper I, 
as well as the 62- and 112-MHz observations by SLS12, clearly 
demonstrate that the MP is a profile in its own right.  Earlier we argued that the MP 
half-power width of 9.5 -- 10\degr\ in both the Q- and B-mode (at around 1 GHz) 
was consistent with B1822--09 having an inner-cone/core triple (T) profile, and its evolution 
at low frequency (see SLS12) is compatible with this expectation.  

Also, we now find that the IP at metre wavelengths is a roughly 15\degr-wide region 
rather than a single feature, with the peak trailing (probably a ``partial cone'' as in ET 
IX terms).  The IP's double structure is clearly confirmed in SLS12's 62-MHz profile (their 
fig. 7), and the overall structure is compatible with an outer conal-double profile, both 
in its 1-GHz dimension of about 13\degr\ (=5.75\degr\ $P_1^{-1/2}$), but also in the 
manner of its frequency evolution.  

Overall, then, we have every indication---on the basis of dimensions, evolution, and 
polarization---that both the MP and IP emission originates from the open dipolar 
field-line regions, which is true for most other radio pulsars.

The origin of B1822--09's IP emission has long been debated.  Early workers had 
interpreted its `main-pulse' as being comprised of the PC and MP together in the 
manner of a conal-double profile (notwithstanding the peculiar behaviour of its PC 
vis-a-vis IP and a raft of other problems with this interpretation).  In these terms, 
the PC-MP centre precedes the IP peak by roughly 186\degr.  

Using, however, the fuller understanding summarized above, the MP centre to IP 
centre separation is 180\degr\ with an error of about $\pm$1\degr---and apparently 
invariant over the entire range of available observations.  This is just the geometry 
for which IP emission from the second pole is expected, and indeed this ``orthogonal 
rotator'' two-pole configuration seems to be provide the best explanation for the IP 
emission.

The properties of the PC then contrast strongly with those of the MP and IP as we 
have seen above, in RF spectrum, polarization, profile evolution, and single-pulse 
properties [\eg Gil \etal\ (1994): fig. 6].  The location of its region of emission is 
currently unknown; however, the B1822--09 PC resembles the similar feature in 
B0943+10 very closely.  

In an effort to explain the remarkable anticorrelation between the star's IP and PC 
emission, Dyks \etal (2005) suggested that the IP and PC emission in fact originate
from the same polar region as the MP, but for some reason the IP and PC emission 
reverse direction during a mode, one emitting in an outward and other in an inward 
direction.  They suggest two kinds of arrangement for this situation.  In the first case 
the IP-PC-MP radiation all comes from the same pole, the PC and MP outwardly 
and the IP representing the inward emission of the PC.  The other configuration has 
the IP and PC arising from the second pole, where the IP represents outward, and 
PC inward, emission. Unfortunately, we doubt that this ``downward emission'' model 
is compatible with observations.  First, the IP and MP are separated by almost exactly 
180\degr, so a very special design is needed to oppose the IP and PC.  Second, the 
PPA traverses below the IP and PC emission are very different (see ET IX); whereas, 
if they arise from the same diverging field-lines they should be similar.  
In any case, the displacement due to aberration/retardation (A/R) of twice 300 km or
so is only about 1\degr, so the spacing cannot decide or rule out this conjecture.  
Thus we believe that the IP and MP emission originate from two magnetic poles and the PC 
emission is on the MP emission side.  

The origin of the PC emission is currently unknown.  Given its proximity to the MP, it 
likely originates above the same pole as the MP and, if associated with the open field 
of the polar flux tube, then it must be emitted higher in the magnetosphere than the MP.  
If, for instance, it originates close to the magnetic axis, then an aberration/retardation 
effect of some 15\degr\ corresponds to an emission height of around 5000 km.  The 
properties of the PC emission are very different from those of the MP or IP emission:  
Both the 50\% width of the PC feature and its separation from the MP are practically 
constant over a wide frequency range from 0.3 MHz to 11 GHz (see Gil \etal\ (1994): 
fig. 6), whereas the IP and MP emission widths decrease by a factor of 2$-$2.5 over 
this band.  The PC emission is also highly linearly polarized (Johnston \etal\ 2008) with 
a flat PA traverse (\eg Backus \etal 2010), suggesting that it may be emitted above the 
polarization limiting region (Johnston \& Weisberg 2006).  Our analysis shows a weak 
fluctuation spectral feature at 0.06 c/$P_1$ for the PC and also ``spiky'' microstructure-like 
emission with time-scales of about 150$\mu$sec as first observed by Gil \etal\ (1994). 
Pre/postcursor emission is only seen in a small group of slower pulsars (see ET IX), 
and does not seem to depend on the pulsar geometry.  

B1822--09 is the only interpulsar known to exhibit profile mode changes, and the near 
anticorrelation of its IP and PC have fascinated pulsarists.  During the pulsar's Q-mode, 
strong, correlated roughly 46-$P_1$ modulation is seen in its IP and MP and the PC is 
nearly silent; whereas, in the B-mode the IP disappears and PC emission precedes the 
MP by some 15\degr, connecting by a low level bridge of emission to the MP.  Our study 
above shows this phenomenon in unprecedented detail.  The overall modal sequencing 
is shown in Fig.~\ref{fig3n}; whereas, the longest respective Q- (7201 pulses) and B-mode 
(1026 pulses) subsequences are depicted in the top and bottom displays of Fig.~\ref{fig1}, 
the first showing the IP and MP and the latter the PC and MP.  

Interestingly, we have found that the two modes are not as completely ``pure'' as formerly 
thought.  We noted earlier that even during the longest Q-mode PS, a few pulses with emission 
at the PC longitude and a weakening IP can be seen.  Or, during the longest B-mode PS, 
one can see IP emission at a very low (1\%) level.  This ``flickering'' between the two 
modes seems to be a general phenomenon, and we see short intervals throughout the 
observation where both the IP and PC are `on'.  Our long sensitive observation allows 
us to more accurately characterize B1822--09's modal phenomenon.  IP emission seems 
to be present during both B- and Q-mode intervals, but is strong and consistent only in 
the Q-mode when the PC is absent; otherwise the IP is extremely weak during the B-mode 
when PC reappears.  Modal transitions seem to require a few pulse periods (\ie 
a few seconds), and we find many instances where all the three emission features are 
simultaneously present for short intervals (\eg see Fig.~\ref{fig2n}).  All this indicates that 
the mode changes entail changes at both the magnetic poles. 

In B1822--09, the strong stable modulation, affecting both the MP and IP, occurs during 
the Q-mode; here the pulsar is somewhat weaker in intensity, but a more ``perfect drifter'' 
(just the opposite as in B0943+10, where the accurate modulation occurs in the B-mode; 
see \eg Backus \etal 2010).  In both pulsars, the appearance of the PC seems to disturb 
the orderly dynamics of the MP; in B0943+10 the highly regular drift becomes chaotic; in 
B1822--09 the regular modulation in the leading and trailing conal sections of the MP 
becomes much less pronounced.  Interestingly, the geometries of the two stars could not 
be more different, in that B0943+10 has a nearly aligned magnetic axis.  

Turning now to the fluctuation spectral properties of PSR B1822--09, we found above 
that the long Q-mode observation exhibited a strong modulation with an apparently 
constant period of 46.55$\pm$0.88 $P_1$.  Further, this fluctuation was seen in both 
the MP and IP.   Both the constant phase of the modulation across the IP region and 
the HRF indicate that this represents an amplitude modulation of the IP.  By contrast, the 
phase under the MP shows a sharp ramp across the pulse, indicating (together with its 
HRF) that the fluctuations are largely phase modulated.  We have also clearly detected 
a weak 70$\pm$3 $P_1$ modulation in the longest B-mode sequence; however, no 
such feature was detected in the other shorter B-mode sequences.  

The Ruderman \& Sutherland (1975) model connects the modulation features of 
``drifting'' subpulses to ``sparks'' undergoing {\bf E}$\times${\bf B} drift while 
breaking down the inner vacuum gap.  The circulation time predicted by this model 
is some 43-$P_1$ which, although close to the observed Q-mode periodicity, should 
be taken as low by a factor as large as 3 or so---because in virtually every instance 
where a CT has been determined, it is much longer than that estimated by RS (\eg 
for B0943+10 the RS-predicted CT is a factor of 3.5 faster than the observed value).  
Thus, we interpret this modulation as a $P_3$ ``drift'', not the CT.  Indeed, in reaction 
to the slower observed CTs, Gil \etal\ (2003) predicted that the inner vacuum gap 
was in fact partially screened by ions, decreasing the electric field and slowing the
circulation.  

 In fact,the CT is related to the parameters of the inner vacuum gap as CT $\sim ~2 a/\eta$, 
where $\eta$ is the screening factor and $a$ is the complexity parameter $a = r_p/h$ 
which is the ratio of the polar cap radius, $r_p$, and the vacuum gap height, $h$, 
(see Eq. 11 of Gil \& Sendyk 2000 and Eq. A1 of Gil \etal\ 2008).  Assuming the 
screening factor\footnote{The cannonical value of $\eta \sim 0.3$ can also be 
understood by analysing the case of B0943+10 where the observed CT is three 
times longer than what is expected from the RS75 vacuum gap model.  In the partially 
screened vacuum gap model this implies that the potential drop across the gap is
three times lower giving $\eta \sim 0.3$.} is $\eta \sim 0.3$ (see Bhattacharyya \etal\ 
2010), and for PSR B0943+10 $a\sim6$ and for PSR B1822$-$09 $a\sim20$, 
the CT is about 40 and 133 pulsar periods respectively, which is close to what is 
observed.  Also recently, van Leeuwen \& Timokhin (2012) have argued that RS's 
calculation of the circulation time was approximate, and if the varying accelerating 
potential across the inner vacuum gap is considered in detail, then the slower CT 
can be explained.  In any case, we have several lines of evidence that the CT may 
be upwards of three times our putative $P_3$ value, and for all the above reasons 
the corresponding 140-$P_1$ is not surprising.  Moreover, the value is commensurate 
with the weak evidence that the B-mode $P_3$ of 70-$P_1$ is half the CT.  This 
interpretation implies a carousel of three MP ``sparks'' in the Q-mode and two sparks 
in the B-mode. 

The existence of the same modulation in B1822--09's IP and MP is puzzling. Similar 
behaviour is also seen in two other interpulsars, B1055--52 and B1709--22 [Weltevrede 
\etal\ (2007, 2012)].  In addition to having the same period, we find a that phase-locked 
relationship exists between the MP and IP modulation.  However, the nature of the 
``locking'' is more complex in the present case than in the stars above as seen, for 
instance, in the modfold plot of Fig.~\ref{fig9}.  Focussing on the IP one can see that the 
intensity drops between pulse number 20 to 40 within the cycle, and the corresponding 
lower intensity for the MP is between about 8 to 28.  This clearly shows that the IP and 
MP pulse sequences are offset in modulation phase by approximately 12 rotation 
periods---a value that is far too long to be a physical delay.  More sensitive observations 
are needed to determine this delay accurately using cross-correlation analysis.  However, 
a phase-locked condition exists between the IP and MP exists for B1822--09 as well as 
the other two interpulsars, signifying some information transfer between the two emission 
regions. 

Additionally, for PSR B1822--09 the appearance of the PC signals both diminished IP 
and disordered MP emission, which is also an indication of information transfer.  Inevitably 
then, we are driven to the conclusion that some kind of global magnetospheric exchange 
exists between the two magnetic pole regions.  The current magnetospheric models of 
radio emission arising within the open dipolar field-line ``flux tubes'' are unable to 
address this circumstance.  In these models the out-flowing particles from the two poles 
stream out past the light cylinder on disconnected regions of open field.  While it is 
possible that the CTs within the two polar regions are similar, since this depends on  
the surface magnetic field and rotation period, the phase-locked condition is indeed 
hard to understand.  Weltevrede \etal\ (2012) discuss various physical models which 
can explain this phenomenon; however, their models are not developed to the extent 
that they can be applied to individual observations. 

The focus of this work has been to characterize the peculiar moding behaviour of pulsar 
B1822--09 using the longest ever (37399-pulse) continuous and sensitive single pulse 
observation from the GMRT.  Our analysis shows that parallel IP and PC emission exist in 
both the pulsar's modes.  The long modal intervals also allow us to accurately determine 
the modulation periods of the IP and MP emission, which we then use to estimate a CT 
for the pulsar.  Finally, we emphasize that the PC emission influences mode changes in 
both this pulsar as well as PSR B0943+10.  It is extremely important to develop theoretical
conceptions regarding the origin of the PC emission.  Overall, our analyses strongly suggest 
that mode changes entail global magnetospheric changes and information transfer between 
the two magnetic polar regions. 

\vspace{5mm}
\noindent{\bf Acknowledgments:}
We sincerely thank our referee Prof. Janusz Gil for providing helpful and 
constructive comments that have improved the paper as well as our collaborators 
Svetlana Suleymanova and Geoffrey Wright for their contributions and/or critical 
readings of earlier versions of the manuscript.  We gratefully acknowledge 
Isaac Backus for assistance with processing the observation, and we thank the 
GMRT staff for providing technical and logistic help during the observations. 
GMRT is operated by the National Centre for Radio Astrophysics of the Tata 
Institute of Fundamental Research.  One of us (JMR) thanks the Anton Pannekoek 
Astronomical Institute for their generous hospitality, the NWO and ASTON for their 
Visitor Grants, and the NCRA for their generous hospitality and registration 
assistance. Portions of this work were carried out with support from US National 
Science Foundation Grants AST 08-07691.  This work used NASA ADS system.

{}

\end{document}